# Selective molecular capture mechanism in carbon nanotube networks

Yu Wan[a], Jun Guan[b], Xudong Yang[b], Quanshui Zheng[a] and Zhiping Xu[a,*]

[a]Applied Mechanics Laboratory, Department of Engineering Mechanics, and Center for Nano and Micro Mechanics, Tsinghua University, Beijing 100084, China

[b]Department of Building Science, School of Architecture, Tsinghua University, Beijing 100084, China

[*]Email: xuzp@tsinghua.edu.cn

**Abstract**

Recent air pollution issues have raised significant attention to develop efficient air filters, and one of the most promising candidates is that enabled by nanofibers. We explore here selective molecular capture mechanism for volatile organic compounds in carbon nanotube networks by performing atomistic simulations. The results are discussed with respect to the two key parameters that define the performance of nanofiltration, i.e. the capture efficiency and flow resistance, which validate the advantage of carbon nanotube networks with high surface-to-volume ratio and atomistically smooth surfaces. We also reveal the important roles of interfacial adhesion and diffusion that govern selective gas transport through the network.

## 1. Introduction

Air pollution, which accompanies the development of human civilization since the first creation of fires, acts as a silent killer to human health.[1] Recent crisis such as the massive haze in China has caused significant economic losses and led to serious social panic. Volatile organic compounds (VOCs) are carbon-rich compounds that participates in atmospheric photochemical reactions, and are also one class of common indoor contaminants.[2] They are emitted from a wide array of products including many building materials, furnishings, office equipments, etc.[2] Long-time exposure to specific VOCs has known healthy effect, e.g. headaches, nausea and serious sensory irritation symptoms, or even correlation with cancer.[3] Effective removal of VOCs from aerogels in the environment thus becomes of paramount importance. This is especially significant for closed environment (e.g. aircraft cabin, vehicles) as recent survey reported that adults spend about 87% and 6% of their time in buildings and vehicles, while only 7% outdoor.[4] More than two million people's deaths have direct or indirect relationship with air pollution each year, and 70% of them are attributed to the indoor air pollution.[5,6]

Efficient techniques are thus urged to maintain the air quality under poor air circulation and diverse sources of pollutants in closed environment. Nanofibrous membrane is one of the most popular materials nowadays used for air filtration, as well as related applications such as water cleaning and food processing. Gas separation by nanofibrous membranes is commonly known to be achieved by molecular sieving assisted by molecular adhesion and diffusion, where molecular transport differs by their masses, sizes, interaction with the fibers, etc.[7,8] Materials of nanofiber or nanosheet networks, e.g. polymers, carbon nanotubes (CNTs), graphene, and nanoporous materials, have attracted much interest recently.[9-14] Nanofiltration using membranes with pore size $d$ from 0.1 to 10 nm feature higher filtration efficiency compared to microfiltration and ultrafiltration techniques. The theory of viscous flow predicts that the pressure drop through a pore scales as $\sim d^{-2}$,[7] and thus the rising energy cost in nanostructured membranes could limit their applications. A recent work reported that functionalization with CNTs increases the specific area of the quartz fiber by 12 times, and significantly improves the filtration efficiency. However, the pressure drop measured only increases slightly.[11] To understand the underlying mechanism and shed light on the design of nanostructured air filters, we explore the filtration process in this work at nanoscale by performing atomistic simulations.

CNT-based fiber network becomes an ideal candidate for filtration and separation applications because of their very high surface-to-volume ratio and atomistically smooth graphitic surface that significantly reduces frictional flow resistance, offering low pressure drop in applications.[10-12, 15, 16] In addition to molecular sieving where gas molecules are separated according to their size, contrastive adhesive strengths to the graphitic wall lead to additional selectivity that is defined by the shape and chemistry of molecules as well. This merit could thus elevate the filtration efficiency without significant cost of mechanical energy against the drag force. The length scale of CNTs and their meshes in networks is down to nanometer, which is even shorter than the mean free paths of gas molecules, which is ~100 nm at ambient condition. The continuum model of gas transport breaks down at this scale and neither the filtration efficiency nor flow resistance could be predicted in this picture. Instead, we take an atomistic approach here based on molecular dynamics (MD) simulations. We will investigate interactions between molecules and CNTs first and then the mechanisms of molecular capture and flow resistance.

Recently we measured VOC concentration in aircraft cabin environments, and reported that toluene ($C_7H_8$), decanal ($C_{10}H_{20}O$), and isobutanol ($C_4H_{10}O$) are at the forefront among the most detected VOCs in more than 100 flights, with detection rate (DR) = 100%, 90% and 64%, respectively.[17] In this study, we choose them and nitrogen ($N_2$) as guest molecules to characterize the gas-CNT interaction and filtration efficiency of a CNT membrane. This choice is also made by considering their distinct molecular structures. Classical MD simulations are performed to simulate the energetics and dynamics of gas separation, using the large-scale atomic/molecular massively parallel simulator (LAMMPS) package.[18] We consider both covalent bonds in VOCs and CNTs and intermolecular non-bonding interactions (electrostatic and van der Waals) in our simulations by using the reactive force field (ReaxFF).[19] Periodic boundary conditions (PBCs) are used in our MD simulations, with a length of 4.9 nm along the CNT axis, and lateral dimensions tuned to define the interwall distance between CNTs. The carbon nanotubes are fixed during the simulations. The flexibility of CNT, although neglected in the current setup, may play more important roles when the mesh size of the CNT network, which corresponds to the simulation box dimension here, increases. However, by assuming a dense mesh of CNTs in the membrane with pore size on the order of a few nanometers, it is reasonable to exclude this effect in this study.

## 2. Results and Discussion

**Interfacial adhesion and sliding strength.** We first quantify interfacial adhesion between guest molecules and the graphitic surface. Equilibrium MD simulations are performed by constraining the relative position between the center of mass of molecule and the underlying graphene lattice, as illustrated in **Fig. 1a**. The results are summarized in **Fig. 2**. The adhesive strengths $\sigma_a$ for nitrogen, toluene, isobutanol and decanal are $1.38\times10^{-10}$, $7.03\times10^{-10}$, $4.12\times10^{-10}$, $7.95\times10^{-10}$ N, and the sliding strengths $\tau_s$ are $0.37\times10^{-11}$, $0.61\times10^{-11}$, $2.89\times10^{-11}$, $1.91\times10^{-11}$ N. These results indicate that the decanal molecule features the highest resistance to pulling off and sliding due to its extended chain-like conformation, followed by toluene consisting of a benzene ring that binds preferentially with the hexagonal graphene lattice. The nitrogen molecules, in contrast, are rather mobile on the surface with low adhesive and shear strengths, which are only 17.4% and 24.7% of the values for decanal. The adhesive energy $E_a$ and energy corrugation $E_s$ upon sliding summarized in **Fig. 2b** show similar trends for these four guest molecules. Compared to the thermal energy $3k_BT/2 = 6.17\times10^{-21}$ J at room temperature, the adhesive energies are 1-2 orders higher, but the energy barriers for diffusion on the graphene lattice is comparable to $3k_BT/2$, suggesting that an adhesion-diffusion mechanism will govern the gas transport process. Similar behavior has been identified for molecular diffusion inside CNTs that enables rapid transport of gases.[15, 20]

**Filtration efficiency.** We now quantify the selectivity by calculating the collection efficiency. To model the network structure of a CNT membrane filter studied in recent work,[10-12] we construct a linear array of (10,10) carbon nanotubes with diameter $D = 0.81$ nm and interwall distance $d$ ranging from 0.3 to 3 nm (**Fig. 1b**). Molecules with velocities $v$ are placed at a distance of 2 nm from the array, where the amplitude and direction of $v$ is initialized according to the Maxwell-Boltzmann distribution at $T = 300$ K. To manifest efficient sampling in velocity, we perform a large ensemble of MD simulations containing 2000 runs for each type of molecules. The simulation time for each run scales inversely with the velocity projected to the direction towards the array and thus some of them could be quite demanded in computation time. We solve this issue by utilizing the volunteer computing grid established at Institute of High Energy Physics of Chinese Academy of Science.

From the simulation results, the molecules passing through, rebounded back from and adhered to the array are counted and summarized in **Fig. 3**. The results show distinctly high selectivity $p$, i.e. $p_{N2/VOCs}$ = +∞, 54.3 and +∞ for toluene, isobutanol and decanal at $d$ = 0.94 nm, and 4.2, 5.2 and 40.5 at $d$ = 2.80 nm. Correspondingly, the collection or capture efficiency can be defined as $\varepsilon = n_c/n_g$, where $n_c$ is the number of molecules captured, and $n_g$ is the number of geometrical incident molecules. The measurements are performed for a specific time interval longer than the timescale for the molecule to diffuse around the CNT. The results show that $\varepsilon$ = 92.2%, 93.8% and 99.2% at $d$ = 2.8 nm. By increasing the pore size $d$, the effect of size exclusion is weakened and molecular capture mechanism is dominated more by the adhesion. As a result, molecules with weak adhesion ($N_2$) pass through the mesh easily while molecules with strong adhesion (VOCs) investigated here are captured with high efficiency. It should be further noted that, the interaction between $N_2$ and CNT is so weak that the $N_2$ molecules can escape from CNTs by environmental perturbation. This means that collection efficiency for nitrogen will be even smaller when long, and thus flexible nanofibers, are used and higher selectivity can be established.

**Flow resistance.** In addition to the selectivity, flow resistance or pressure drop across the membrane it results is another key indicator to measure performance of a filter. The resistance exerted by the filter onto the molecular flow can be decomposed to momentum and frictional parts that are directly related to the adhesive and shear strengths we have discussed. We measure the momentum drag force by injecting molecules towards a graphene sheet, as the frictional contribution is already indicated as $\tau_s$ in **Fig. 2b**. We find that the reaction force $f_R$ increases with the incident speed $v$ in the range from 10 to 600 m/s explored in our MD simulations. For molecules with a specific kinetic energy of $3k_BT/2$ at $T$ = 300 K, we plot the time evolution of $f_R$ in **Fig. 4**. The results show that $f_R$ for nitrogen is lower than that for VOCs, as can be explained by its weak adhesion to the graphitic surface. Moreover, there are multiple peaks identified for the VOCs, which captures the adhesion-diffusion behavior of VOC molecules on graphene. The corrugation in the profile of $f_R$ corresponds to momentum transfer events between the molecule and graphene due to thermal motion of the molecules.

To quantify the overall resistance on a nanofiber network in the flow, one needs some knowledge of the streamlines. The nature of flow around a fiber can be interpreted by according to the Knudsen number Kn = $\lambda/l$,[8] where $\lambda \sim 100$ nm is the mean free path of molecules and $l$ is the characteristic length scale in the network that is ~1 nm. These are four distinct flow regimes, including continuum or viscous regime (Kn < $10^{-3}$, **Fig. 5a**), slip-flow regime ($10^{-3}$ < Kn < 0.25), free molecule regime (Kn > 10) and a transient regime (0.25 < Kn < 10). However, in comparison with free molecule flow of ideal gas (**Fig. 5b**), the situation here (Kn > 10) is complicated by the facts of adhesion and surface diffusion, as can be seen from a typical molecular trajectory (**Fig. 5c**). The molecule injected is either rebounded back or trapped by the CNT after its first collision with the CNT. For the latter situation, the molecule then diffuses on the surface with successive collisions governed by its normal momentum of thermal motion. It may eventually escape from the CNT at time and location that is controlled by the thermodynamic equilibrium fluctuation and cannot be well predicted. This effect should be considered for quantitative prediction of the resistance, for example by including a 'diffuse reflection' assumption where the velocity of detached molecules is defined by equilibrium thermodynamics.[21] However, it should be noted that the retardance due to surface diffusion of the molecules plays an additional role in modulating the resistance.

Consider a molecule moving toward a nanofiber, the velocity $v$ can be decomposed into its normal and tangential components $v_n$ and $v_t$ before it collides with the fiber. The amplitude of pre-collision velocity $v_n$ determines the behavior of collision. The velocity right after the first collision $v_{n1}$ usually increases with $v_n$ and is lower than $v_n$ due to the energy loss into thermal vibration. We define a critical normal velocity $v_{nc} = (2E_a/m)^{1/2}$ from the adhesive energy $E_a$. The molecule then will be captured by the nanofiber as $v_{n1}$ < $v_{nc}$, or be able to escape for $v_{n1}$ > $v_{nc}$. After being captured, the incident kinetic energy decays and the normal velocity of molecules then will be governed by thermal fluctuation. The captured molecule may eventually escape from the nanofiber when its instantaneous normal velocity exceeds $v_{nc}$ after a certain times of collisions.

To obtain some insights into the momentum and energy transfer between the molecules and CNTs, we further define the coefficient of restitution as $α_1 = -v_{n1}/v_n$. For molecules with specific kinetic energy of $3k_BT/2$ at $T$ = 300 K, the results show that $α_1$ = 0.95, 0.29, 0.65, 0.28 for nitrogen, toluene, isobutanol and decanal, indicating a significantly higher energy loss due to stronger adhesion between VOCs and graphene than nitrogen. Additional energy channeling exists as well from the normal motion to the tangential ones. The flow resistance contributed by the momentum transfer is then dominantly determined by the first few collisions and can be reduced by considering diffusive motion of the molecules at this scale where correlation between independence collision events is lost, in contrast to ideal gas or viscous flow as shown in **Fig. 5**. Additionally, according to the measured values in **Fig. 2**, frictional drag force on the graphitic walls is negligible ($E_s$ ~ $3k_BT/2$). This may explain the low pressure drop across CNT-based filters recently reported.[11]

## 3. Conclusion

In summary, we identify the molecular mechanism to achieve high selectivity and low flow resistance in CNT-based nanofiber network as efficient nanofiltration media for VOCs. The selective gas transport involves distinct adhesion-diffusion processes and our atomistic simulation results show that the measured extraordinary performance is attributed to the contrastive adhesive strength with CNTs, between VOCs and molecules such as nitrogen. The diffusive nature of molecular motion at nanoscale and low friction on the graphitic surface reduces the resistance against molecular flow. These findings lay the ground for optimal design of nanofiber-based filtration materials for clean air applications.

**Acknowledgment**

This work was supported by the Boeing Company, the National Natural Science Foundation of China through Grant 11222217, 11002079, Tsinghua University Initiative Scientific Research Program 2011Z02174. This work was performed on the volunteer computing grid CAS@Home (http://casathome.ihep.ac.cn/), and the Explorer 100 cluster system of Tsinghua National Laboratory for Information Science and Technology.

**FIGURES AND CAPTIONS**

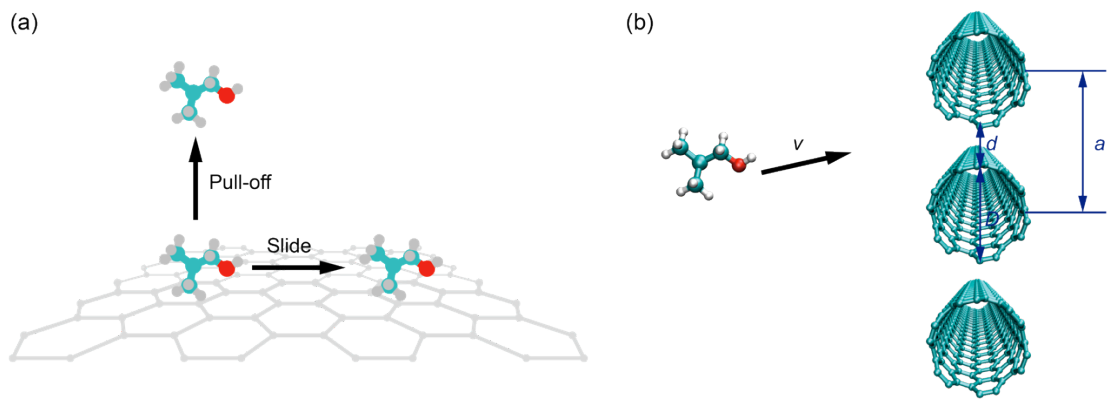

**FIG. 1.** MD simulation setups for measurements of (a) adhesive and shear strengths between molecules and a graphitic surface, and (b) filtration performance of a linear CNT array.

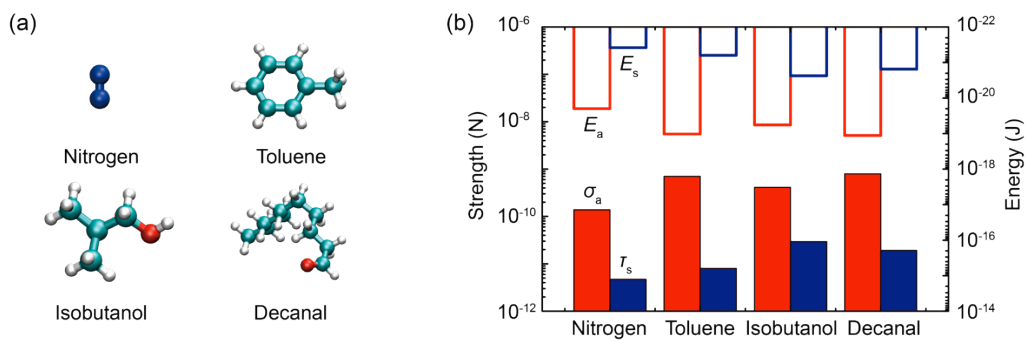

**FIG. 2.** (a) Molecular structures of nitrogen and VOCs explored in this work. (b) Adhesive strength $\sigma_a$ and shear strength $\tau_s$ (filled bars), adhesive energy $E_a$ and energy corrugation along sliding $E_s$ (open bars).

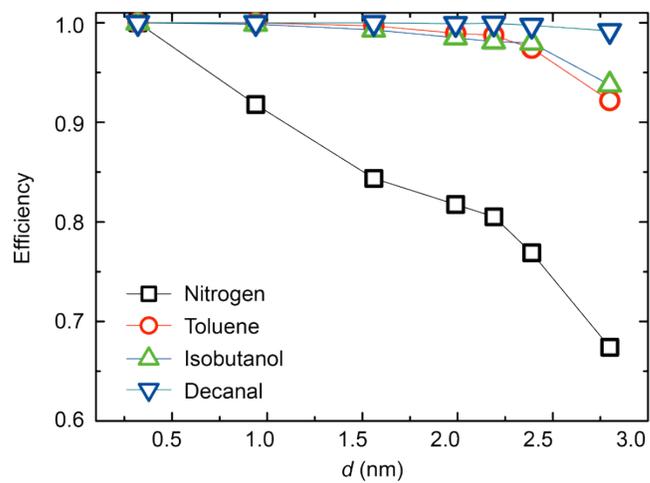

**FIG. 3.** Molecular capture efficiency for nitrogen and VOCs, measured from MD simulations.

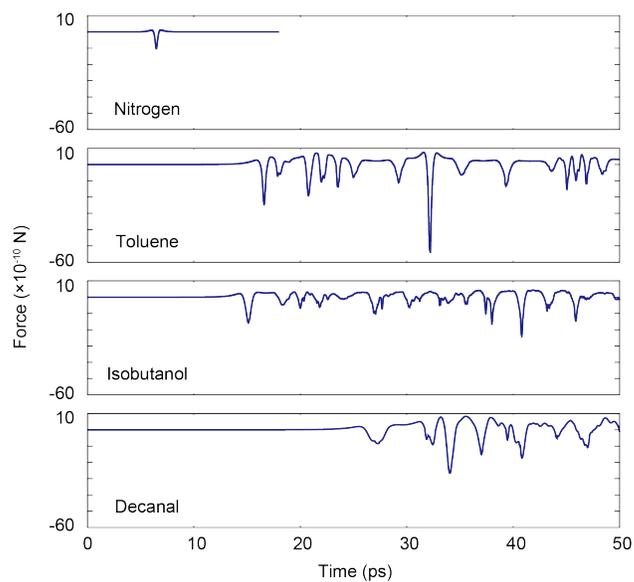

**FIG. 4.** Forces experienced by the in-flow molecules moving towards a graphene sheet. Multiple peaks in the force evolution for VOCs indicate the adhesion-diffusion mechanism of molecules on graphene.

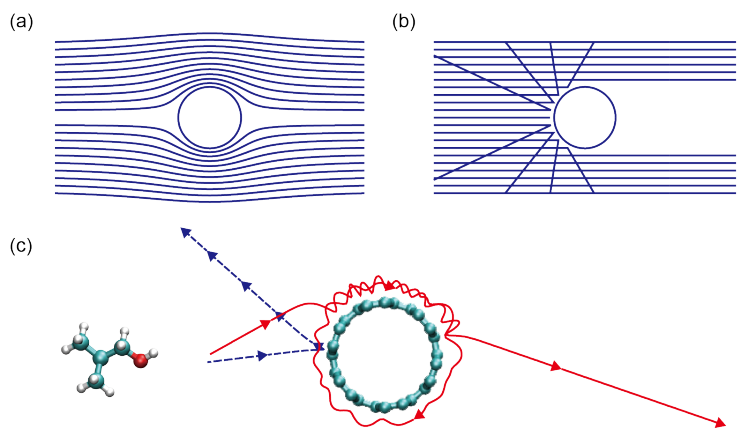

**FIG. 5.** Streamlines for (a) viscous and (b) ideal gas flow against a cylindrical obstacle. (c) MD simulated trajectories of molecules moving toward a CNT, showing distinct behaviors for captured (red, solid line) and rebounded (blue, dash line) states after collision with the CNT wall.